\documentclass[aps,amsmath,amssymb, twocolumn]{revtex4-1}
\usepackage{amssymb, amsmath}
\usepackage{bm}
\usepackage{graphicx}
\usepackage{color}
\usepackage{amsfonts}
\usepackage{latexsym}
\usepackage{subfigure}
\newcommand{\hh }[1]{ \hat{\bm{#1}} }
\newcommand{\m }[1]{ \mathbf{#1} }

\def\eq#1{{Eq.(\ref{#1})}}    \def\fig#1{{Fig.\ref{#1}}}

\def\sec#1{{Section \ref{#1}}}
\def\app#1{{Appendix \ref{#1}}}

\usepackage{hyperref}

\hypersetup{
 linktocpage=true,
 pdfborderstyle={/S/S/W 1},
 hyperindex=true,
 bookmarks=true,
 bookmarksopen=true,
 bookmarksnumbered=true,
}

\begin{document}

\title{
Polarisation of cells and soft objects driven by mechanical interactions: \\  Consequences for migration and chemotaxis  }

\author{M. Leoni \&  P. Sens}
\affiliation{Laboratoire Gulliver, UMR 7083 CNRS-ESPCI, 10 rue Vauquelin, 75231 Paris Cedex 05 - France}

\begin{abstract}
  We study a generic model for the polarisation and motility of  
  self-propelled soft objects,  biological cells or
    biomimetic systems, interacting with a viscous substrate.
 The active forces generated by the cell on the substrate are modelled 
by means of oscillating force multipoles at the cell-substrate interface. 
Symmetry breaking and cell polarisation  for a  range of cell sizes 
 naturally ``emerge'' from long range mechanical interactions between oscillating units, mediated both by the intracellular medium and the substrate. However, the harnessing of cell polarisation for motility requires substrate-mediated interactions.
 Motility can be optimised by  adapting the oscillation frequency to the relaxation time of the system  or  when the substrate and cell viscosities match.   
Cellular noise can destroy mechanical coordination between force-generating elements within the cell, resulting in sudden changes of  polarisation. The persistence of the cell's motion is found to depend on  the cell size and the substrate viscosity. Within such a model, chemotactic guidance of cell motion is obtained by directionally modulating the persistence of motion, rather than by modulating 
the instantaneous cell velocity,
 in a way that resembles the run and tumble chemotaxis of bacteria.
   \end{abstract}

\date{\today}
\maketitle

\section{Introduction}

Cell motility on solid substrates (crawling) and in fluid environments (swimming) are usually regarded as being based on different physical principles and studied independently. 
Swimming
often relies on the beating or rotation of protrusive appendages (flagella or cilia)~\cite{Bray,Lauga},
 while crawling generally relies on the protrusive and contractile forces generated by the acto-myosin cytoskeleton, and on cell-substrate adhesion~\cite{Bray}. 
Remarkably, several cell types that undergo major shape changes while crawling, such as amoebae and neutrophils, are 
 also
 able to swim 
in Newtonian fluids~\cite{Barry,Barry2}. Furthermore, crawling cells are sensitive to the stiffness (elastic response)~\cite{gardel:2008,Stroka,Pathak}, but also to the viscosity~\cite{Kourouklis,Muller} of the substrate on which they are crawling. 
This suggests that crawling and swimming share common underlying physical principles and that insight on eukaryotic cell motility may be gained 
 by studying self-propelling  soft objects in fluid environments~\cite{Fartuin}.

Cell motility requires polarisation, namely the breaking of front-back  symmetry of the cell.
  This can be triggered by external gradients of either biochemical (chemotaxis)~\cite{chemotaxis-rev10} or mechanical ({\em e.g} durotaxis)~\cite{durotaxis} nature, or it can occur spontaneously. Spontaneous cell polarisation is of particular interest to understand the spatio-temporal correlations inside and outside the cell.  Active gel theories~\cite{Sykes,Sekimoto04,Kruse,Ziebert,Tjhung,Kozolov} have shed some light on the physics of spontaneous cell polarisation, but much remains to be learnt about how coordination is achieved at the  scale of the whole cell,  and  how   coordinated motions are affected by the properties of the extra-cellular environment. 
 
Here, we 
devise a generic theoretical framework which enables us to address in a unified manner three different fundamental aspects: polarisation, motility and chemotaxis. Our model soft cell (henceforth simply `cell'), which aims at describing active droplets~\cite{Kruse,Ziebert,Tjhung,JR12}, bio-mimetic self-propelled systems and cells, is a viscous or an elastic body interacting with a substrate by means of localised forces distributed over the cell-substrate contact area. 
We restrict ourselves to the case where the substrate is a Newtonian  fluid, so that force transmission between the cell and the substrate can be  achieved by enforcing a no-slip boundary condition. 
Our findings  also  provide   insight into the  polarisation of cells adhering  to elastic substrates, although 
crawling in that  case  requires additional assumptions concerning the dynamic of creation and destruction of adhesion sites. 

The cell cytoskeleton  commonly displays quasi-periodic spatio-temporal patterns and fluctuations  of activity~\cite{Killich,Ji,Graziano}. 
These  oscillations are observed  in both  adherent~\cite{Ghassemi,Ladoux,schwarz:2013} and crawling cells~\cite{Tanimoto14,galbraith:1997,giannone:2004}  
even in  the absence of specific adhesion receptors~\cite{Lammermann08}.  
Motivated by that, we model the distribution of cellular forces  
as oscillating force-multipoles  at the cell-substrate interface. 
Spatio-temporal correlations among the oscillating forces
  may spontaneously emerge from mechanical interactions,
    affecting cell polarisation and cell motility.
To show this, we first study the motility of a cell exerting forces on the substrate with prescribed time-dependence. 
 The  resulting cell speed has non-monotonic trend as function of the substrate viscosity  or of the frequency of the oscillating units.
We then study spontaneous
 symmetry breaking associated to cell polarisation and the transition from non-motile to motile states
  as a dynamical process emerging 
 from the phase-locking of the oscillating elements. This process  
occurs only  for a range of  cell sizes and is  permitted   
   by long-range interactions mediated by both the substrate and the cytoplasm.
 To investigate chemotaxis, 
we introduce a minimal coupling between chemotactic gradients
and the force distribution. Finally, we discuss the persistence of cell motion in the presence of 
 noise (e.g. cellular noise) and we obtain 
 testable  predictions regarding the  substrate-dependent statistical  properties of the cell trajectories.

\begin{figure}[t]
 \centering
    \includegraphics[width=0.47\textwidth]{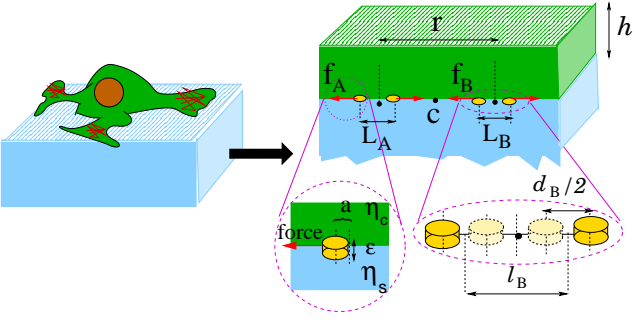}
    \caption{ (Color online)
A  motile cell exerting traction forces on a flat viscous substrate is modelled as  two coupled fluids: a layer of thickness $\mathfrak{h}$ and viscosity $\eta_c$  (the ``cell'' on top, in green) lying over an infinite bulk fluid substrate of viscosity $\eta_s$ (at the bottom, in light blue). Cytoskeletal elements generating traction   are modelled as oscillating force multipoles  (two dipoles in the sketch) distributed along a line. }
  \label{sketch}
 \end{figure}  
 
 \section{Model}
To begin with, we  model the cell as  an incompressible fluid layer of  thickness $\mathfrak{h}$ and viscosity $\eta_c$ lying over an  incompressible bulk fluid of viscosity $\eta_s$  (\fig{sketch}) at low Reynolds number. The fluids are immiscible and  satisfy  the Stokes equation~\cite{Landau6}:
\begin{align}
 & \eta_c \nabla^2 \m v^{(c)} -\bm{\nabla} p^{(c)}  = 0 
 ; \bm{\nabla} \cdot  \m v^{(c)} = 0, (0 \le z \le \mathfrak{h}) \label{eq:coupled-fluids} \\
 & \eta_s \nabla^2 \m v^{(s)}  -\bm{\nabla} p^{(s)}  = 0 
 ; \bm{\nabla} \cdot  \m v^{(s)}  = 0, 
(-\infty<  z < 0 )\nonumber
\end{align}
where  $\m v^{(c)}, p^{(c)}$ and $\m v^{(s)}, p^{(s)}$ are the velocity  and pressure fields  in the fluid layer and  the substrate, and  $z$ the direction normal to the interface. The fields $\m v^{(c)}$ and  $\m v^{(s)}$ are  coupled  by means of conditions at  the interface $z=0$.  In the following, we assume no-slip boundary conditions at the interface, $\m v^{(c)}{}_{|_{z=0}}=\m v^{(s)}{}_{|_{z=0}}$, and  neglect  normal displacements of the interface, $v^{(c)}_z{}_{|_{z=0}}=v^{(s)}_z{}_{|_{z=0}}=0$.  
Similar results are obtained if   no-slip  is enforced only at the location of the active force generators, see~\app{sec:narrow-gap}.

Protein complexes responsible for the traction forces at the cell-substrate interface are modelled as $N$ disks of finite radius $a$ and negligible height $\epsilon\rightarrow 0$ distributed at positions $\m r_n$ ($n=1, \ldots N$) at the interface between the two fluids via  $ \bm{F}(\m r, z=0) = \sum^N_{n=1}   \m F_n \delta(\m r -\m r_n)$. By virtue of Newton's third law, the cell exerts no net force on the substrate, $\sum^N_{n=1}   \m F_n=0$, and the local  stress balance at the interface  reads
 \begin{equation}
\eta_c \frac{d}{dz} \m v^{(c)}_\parallel -  \eta_s \frac{d}{dz} \m v^{(s)}_\parallel =- \bm{F}(\m r, z=0) 
\label{bondcond}
\end{equation} 
where $\m v^{(\varsigma)}_{\parallel}$ denotes the in-plane component of the velocity $\m v^{(\varsigma)}$ for $\varsigma = s, c$, see~\app{sec:two-fluid-layers}.
The  solution of Eqs.(\ref{eq:coupled-fluids},\ref{bondcond}) can be obtained using Fourier transforms, see   e.g.~\cite{Lubensky,Lenz}, as explained in~\app{sec:two-fluid-layers}. 
We focus here on the analytically tractable limit of  two coupled semi-infinite  fluids, $\mathfrak{h} \to \infty$. The interfacial velocity field at a distance $r$ from a unique interfacial point-force $\m f $  is   $ \m v(r) = \frac{1}{2 \pi (\eta_s +\eta_c)} \frac{1 }{r}\m f$. 
 The effect of several point-forces is  additive thanks to the linearity of~\eq{eq:coupled-fluids}.

To study the emergence of cell polarisation we consider 
 time-dependent and  periodic
 force distributions consisting of two identical units $A$ and $B$,  
  labelled with index $\alpha$ in the following,
 that   on average 
 over a period are 
  mirror symmetric with respect to the cell centre (\fig{sketch}). For simplicity, all forces are distributed along a line. Each unit is made of particles with  coordinates $\m x^\alpha_n =  x^\alpha_n \hh x$, subjected to forces  $\m F^\alpha_n = F^\alpha_n \hh x$ directed along  the $\hh x $ axis. Each unit centre is at $\m c_\alpha=2/N \sum_{n} \m x^\alpha_n $, the cell center at $\m c =(\m c_A+\m c_B)/2$, and the
 separation between units is $ \m r := \m c_B-\m c_A$.

\section{Substrate viscosity affects the cell speed}

\subsection{Derivation of the cell speed}

Interactions between the point forces may lead to the motion of the force distribution with respect to both the cytoplasm and the substrate.  
Only the interactions mediated by the substrate can  lead to displacement of the whole cell. Since the cell's radial boundary is not introduced explicitly in our model, the  cell net velocity is obtained by subtracting the cytosol motion to the motion of the force centre $\m c$. This leads to a force balance 
$\zeta \dot{\m c} = \sum_\alpha \sum_n \bm{\mathcal{S}}^\alpha_n $ as explained in detail in~\app{sec:crawl-eq}. Here,  $\bm{\mathcal{S}}^\alpha_n = \zeta_s \m v(\m x^\alpha_n) $ represents the substrate-mediated traction force, 
and
 $\m v(\m x^\alpha_n)  := \m v^{(c)}(\m x^\alpha_n)  = \m v^{(s)}(\m x^\alpha_n)$ 
 is
 the velocity at the location of disk $(\alpha,n)$ due to the motion of all remaining disks. 
 The total and substrate-related drag coefficients are
 $\zeta$ and $\zeta_s$.  In the limit  $\mathfrak{h} \to \infty$ for  a disk of radius $a$ :  $\zeta_s = \frac{16}{3} \eta_s  a $ and $\zeta = \frac{16}{3} (\eta_s +\eta_c) a $~\cite{Ranger}. 

We  first consider the case where each unit is an oscillating force dipole made of two particles: $n = 1,2$ with one force scale $ F^\alpha_1 =  - F^\alpha_2 := f_\alpha$ and one length scale $ L_\alpha =  x^\alpha_{1} -  x^\alpha_{2}$.  With no loss of generality we write $L_\alpha = l_\alpha + d_\alpha$ where $l_\alpha $ is a constant and $d_\alpha$ describes time-dependent deformations, see \fig{sketch}. We parametrize the amplitude and force of the oscillating dipoles as:
\begin{align}
& d_\alpha =  R_\alpha \cos(\omega t +\phi_\alpha) ;
\left. \right.  f_\alpha = - g_\alpha \sin(\omega t +\phi_\alpha).
\label{eq:param-d-f}
\end{align} 
The average cell velocity is obtained from the net displacement of the cell centre over an oscillation period.
For the pair of identical dipoles  ($\{g,R,l\}_A=\{g,l,R\}_B$) shown in \fig{sketch}, for  $\mathfrak{h} \to \infty$   and to lowest order 
in $l_\alpha/r$,  it has a simple expression derived in~\app{sec:mig-speed}: 
 \begin{equation}
v_c (\psi):=\frac{1}{T} \int^T_0 dt \dot{c}(t)  \sim 
\frac{\eta_s}{ (\eta_s +\eta_c)^2}\frac{gRl^2}{r^4}
\Xi
  \sin\psi \label{eq:c-dot-av} 
 \end{equation}
where $\psi := \phi_B -\phi_A$ is the phase difference between the oscillators. Geometrical details of the force distribution are contained in $\Xi=  \frac{3}{32 \pi} [4+  (R/l)^2 \cos  \psi] $, see~\app{sec:viscous-cell}. 
Replacing the viscous  cell with an elastic cell of elastic modulus $\mu_c$ yields a similar result, 
with $\mu_c/(i \omega)$ in place of $\eta_c$, see~\app{sec:elastic-cell},~\app{sec:elastic-cell2},~\app{sec:elastic-cell3}.   

The relationship between forces and displacements is obtained from the force-balance equation: $\zeta  \dot{\m d}_\alpha  = 2 \m f_\alpha + \bm{\mathcal{I}}_\alpha $. Here
$\bm{\mathcal{I}}_\alpha  =\zeta[ \m v(\m x^\alpha_1 ) - \m v(\m x^\alpha_2 ) ]$ represent non-local interactions, 
propagated both by intracellular and extracellular  media. For small amplitude oscillations: $R_\alpha \ll l_\alpha< r$, interactions may be neglected to lowest order and the force-amplitude relation is $g_\alpha=(\zeta\omega R_\alpha)/2$. 

\begin{figure}[t]
 \includegraphics[scale=0.43]{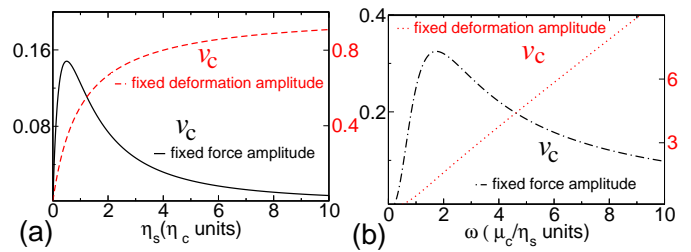}
 \caption{
 (Color online)
 (a) Average migration speed (\eq{eq:c-dot-av})  as function of the substrate 
viscosity $\eta_s$.  The force distribution is shown in \fig{sketch}, with $\psi = \pi/2$, for either fixed force (continuos black  line) or fixed displacement (dashed red  line). 
  The former is measured in units of $(9g^2l^2)/(64 \pi  \eta^2_c\omega a r^4)$ and the latter in units of 
  $ (a \omega  l^2 R^2)/(\pi r^4)$. $\eta_s$ is measured in units of $\eta_c$.
 (b)  Average migration speed variation with the driving frequency for an elastic cell of elastic modulus $\mu_c$. 
  Velocities for fixed force (dashed dotted black  line) and fixed displacement (dotted red  line) are 
  measured respectively in units of $ (9  g^2 l^2)/(64 \pi \eta_s \mu_c a r^4)$ and in units of $(a \mu_c R^2 l^2)/ (\eta_s \pi r^4) $ 
  while  $\omega$ is measured in units of  $\mu_c/\eta_s$, the viscoelastic frequency.
  }
   \label{fig:trend}
 \end{figure}

The active  forces (typically actin polymerisation and actomyosin contraction for crawling cells)  may impose either the force  or the displacement scale. \eq{eq:c-dot-av} shows that  the migration speed  is qualitatively different for imposed displacement ($v_c\sim\eta_s/(\eta_s+\eta_c)\times a\omega(Rl)^2/r^4$) and imposed force ($v_c\sim\eta_s/(\eta_s+\eta_c)^3\times(gl)^2/(a\omega r^4)$). In the latter case, the velocity 
 presents a maximum when the substrate and cell viscosity are similar  (see \fig{fig:trend}(a) - the value of the optimal ratio depends on   $\psi$). This biphasic behaviour shows an interesting analogy with the biphasic velocity of cells crawling on elastic substrates in response to substrate stiffness~\cite{gardel:2008,Stroka,Pathak}.  The speed of an elastic cell can be optimised by tuning the frequency of the oscillating force units  close to the inverse relaxation time of the system:  $\omega \sim 2 \frac{\mu_c}{\eta_s} $, see \fig{fig:trend}(b). This
 suggests that cells may adjust their oscillations rate to the mechanical properties of the environment for optimal motility.
 By symmetry,  a fluid cell  with oscillating disks permanently bound to an elastic substrate experiences a net cytoplasmic flow.  However, harnessing this flow for cell migration on an elastic substrate requires
 additional
  hypothesis concerning disk attachment to and detachment from the substrate, or disk creation at the cell front and destruction at the rear.

   Using typical values of  parameters obtained from experiments
 we estimate   $v_c \sim  10 (\mu m)/hr$.
This is smaller than typical cellular speed,  a common drawback  for swimmers, made up of point-like particles, undergoing small amplitude strokes~\cite{golestanian:2008}. Higher speeds can be obtained by increasing  $a/r$ and $l_\alpha/r$ beyond the validity of the analytical results, \eq{eq:c-dot-av}. This expression is however very valuable, as it shows that net migration requires that the oscillators  {\it phase lock} at  $\psi\ne0(\pi)$,  like 
 interacting dumb-bells in bulk fluids~\cite{LB08,AY08}.
 This breaks  time-reversal symmetry,  a requirement for motility at low Reynolds number
 (``scallop theorem''~\citep{Purcell}).
 
  \subsection{Estimated value for the average migration speed}
Probing the cell with
periodic  stress  or 
strain
of well defined
   frequency $\omega$   
allows one to infer 
  the viscosity  of the cytosol
     $\eta_{c}$, 
   from measured values of the shear  loss modulus $ \mathtt{G}^{''}_{c}$ by means of the relation $\eta_{c} 
   \sim  \mathtt{G}^{''}_{c}/\omega $.
   
To  estimate values of migration speed in our model,
we consider 
oscillation  frequencies 
$\nu_0  \sim 1 Hz $, (a lower bound for swimming cells~\cite{Lauga})  
i.e. 
$\omega  = 2 \pi  (rad)/s$. 
 Intra-cellular
measurements give $ \mathtt{G}^{''}_c \sim 100 Pa $ at
$\nu_0  \sim 1 Hz $, see~\cite{Fabry}. 
  The  associated cell viscosity is  
$  \mathtt{G}^{''}_{c}/\omega = \frac{100}{2\pi} Pa \hspace{0.1cm} s$ (a highly 
 viscous regime:  for comparison water at room temperature has a viscosity $\sim 10^{-3} Pa \hspace{0.1cm} s$ -- four orders of magnitude below).
For simplicity we take a substrate with equivalent viscosity, i.e. we consider conditions of `viscosity matching' between the cell and the substrate.

We next address  the  relation among force and deformation
 for the oscillator. We consider forces
  $g \sim 0.5 n N$
  that are varying with 
  frequency
$\omega  \sim  2 \pi  (rad)/s$
dragging a disk of size $0.5 \mu m$. 
The drag coefficient is 
$\zeta = [ 16 a (\eta_s +\eta_c ) ]/3 
  \sim ( 10 a \mathtt{G}^{''}_c) /\omega  $.
The corresponding oscillation amplitude is
$ R =
\frac{ 2 g  }{\omega \zeta  }
\sim  \frac{10^{-9} N }{10 \times 0.5  \mu m  100  Pa }
\sim 
  2 \mu m .
 $

The average migration speed
 can be calculated using the formula
  valid for fixed deformation amplitudes, reported above.
 In addition to the previous values,
 we consider
$l \sim 2 R \sim  4 \mu m$ and $r \sim 10 \mu m$
(as well as `viscosity matching'  $\eta_s =\eta_c$)
from which 
$ 
v_c \sim 
\frac{a}{\pi}
  \frac{  R^2   l^2\omega  }{  r^4} 
\frac{\eta_s}{( \eta_s + \eta_c)}
\sim (10  \mu m)/hr.
 $
Under  same conditions of
 lengths, and matching between cell-substrate moduli 
$\eta_s \sim \frac{\mu_c}{\omega}$ ,
 the elastic cell has the same speed of the viscous cell
 $v_c\sim  \frac{a}{ \pi} \frac{ R^2 l^2 \omega^3}{ r^4}
 \frac{\eta^2_s}{ (\omega^2 \eta^2_s + \mu^2_c) }
  \sim  (10  \mu m)/hr.
   $

\section{Cell polarisation, motility and chemotaxis emerge from synchronisation}
We now show that  
 polarisation and migration of the cell may spontaneously emerge as the result of dynamical  interactions and synchronisation of the force generators.  The amplitudes $R_\alpha(t)$ and $g_\alpha(t)$,  and the phase $\phi_\alpha(t)$ are now  slowly varying quantities :  constant 
  over an  oscillation period $T=2\pi/\omega$, but  varying over longer time scales as a result of intracellular interactions~\cite{Pikovsky, LLPRE2012}.  
 A simple model~\cite{LLPRE2012} leading to self-sustained oscillations 
 is  the evolution equation for the forces,
\begin{equation}
\dot{f}_\alpha = 
\frac{1}{2}[-\mathcal{K}_\alpha d_\alpha + \mathcal{M} f_\alpha [1 -\sigma d^2_\alpha ] +
\mathcal{A} d^3_\alpha]
\label{eq:f-dot}
\end{equation}
  combined with the force balance equation $\zeta\dot{\m d}_\alpha = 2\m f_\alpha + \bm{\mathcal{I} }_\alpha$  (for $\alpha =A, B$). These equations are
equivalent to  a van der Pol-Duffing oscillator~\cite{Guckenheimer:2002,LLPRE2012}
that exhibits super-critical
  Hopf's bifurcation~\cite{Guckenheimer:2002} in a wide  range of the parameter space. 
 Similar  models  emerge from the collective dynamics of molecular motors, see e.g.~\citep{JP97,Guerin}.
  Here, $\mathcal{K}_\alpha$ sets the oscillation frequency, $\mathcal{M}$ determines  the instability threshold and   $\sigma$ is  a stabilising term. These parameters must be strictly positive for a stable limit cycle to exist. 
$\mathcal{A} \neq 0 $  describes  non-isochronous  oscillations~\cite{LLPRE2012}, and determines whether the oscillation frequency  increases ($\mathcal{A} < 0$) or decreases ($\mathcal{A}>0$) with increasing  amplitude $R_\alpha$. This non-linear coupling may for instance result from the mechano-sensitive kinetics of  sub-cellular constituents.
 
 In the absence of chemotactic bias, the parameters $\mathcal{K}$, $\sigma$ and $\mathcal{A}$ are identical at both ends $A$ and $B$ of the cell. We implement chemotaxis by writing $\mathcal{K}_\alpha  =\mathcal{K} + \tilde{\mathcal{K}} \rho_\alpha $ where  $\rho_\alpha =\rho(c_\alpha)$ is the density of chemoattractant at the centre $c_\alpha$ of unit $\alpha$. This description is consistent with recent studies on chemotaxis~\cite{Insall},  where  pseudopods at the cell edges display wave-like patterns and chemoattractant  changes the rate of internal processes (here the oscillation rate, see below).

The dynamics of a single oscillator is seen neglecting  interactions $\bm{\mathcal{I} }_\alpha$. Combining Eqs.(\ref{eq:param-d-f},\ref{eq:f-dot}) with $\zeta \dot{d}_\alpha = 2 f_\alpha$, and  averaging over the (fast) period $T$ yields $\dot{R}_\alpha = \mathcal{M}  R_\alpha [1 - \sigma (R_\alpha)^2/4]$, showing that the amplitude saturates at a stable value $\mathfrak{R} = 2( \sqrt{\sigma})^{-1} $, and  the phase rotates at slow frequency $\dot{\phi}_\alpha =   -\mathcal{K}_\alpha +    \mathcal{A} \mathfrak{R}^2$. 
To study  synchronisation of  distant oscillators, we assume that the
 interactions
 $\bm{\mathcal{I} }_\alpha$
  induce small deviations from the limit cycle ~\cite{LLPRE2012},  see~\cite{SI} for details, yielding a  phase equation:
\begin{equation}
 \dot{\psi } =  \Omega  + {\rm sgn}[\mathcal{A} ] \frac{1}{\tau_0} (U +  \cos \psi) \sin \psi. 
\label{eq:psidot}
\end{equation} 
Here $\Omega  =\frac{\tilde{\mathcal{K}} }{2 \omega  \zeta } (\rho^A- \rho^B) $ is the  chemotactic bias, the relaxation time $\tau_0$ results from the long-range interactions between oscillators, and $U$  contains geometrical details of the force distribution.
\eq{eq:psidot} resembles  Adler's equation~\cite{Pikovsky}, previously used to describe synchronisation~\cite{Gold,LLPRE2012} 
 and swimming of algae having intrinsic front-back asymmetry~\cite{FJ12,BG13} in low Reynolds number fluids. The additional term $\sin \psi \cos \psi$  is associated to force multipoles. For two dipolar units (\fig{sketch}); $\tau_0=(\pi \mathcal{M} \zeta r^3) /(2 | \mathcal{A}  | \mathfrak{R}^4 a)$ and $U=2l^2/\mathfrak{R}^2$.

\begin{figure}[t]
\includegraphics[width=0.48\textwidth]{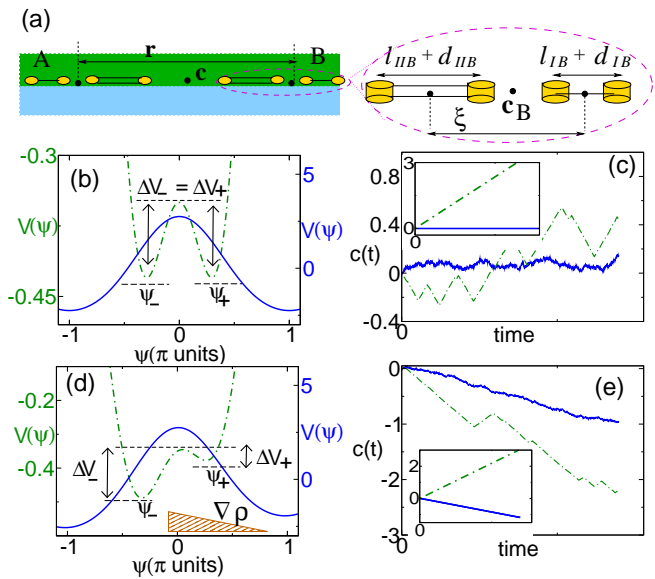}
\caption{ 
(Color online)
(a)  Force-distribution corresponding to two polarised oscillators at either edge of the cell, a system that can show spontaneous symmetry breaking ($|U|< 1$). (b-e) Synchronisation landscape and cell trajectories  obtained from numerical simulations (see~\cite{SI} for details). All panels have $\mathcal{A} > 0$ and either  $U > 1$ (continuous blue curves) or $|U|<1$ (dashed-dotted green  curves). (b) Effective cell polarisation landscape $V(\psi)$, and (c) trajectories of the cell centre $c(t)$ with time in the absence of chemotactic gradient ($\Omega =0$). In the presence of chemotactic gradient, (d), the effective landscape $V(\psi)$ is asymmetric and (e) the cell trajectories show a bias.  Insets in (c,e)  are trajectories in the absence of noise. In this case, the direction of motion for $ |U | < 1$   is arbitrary, and may be away from the chemotactic gradient. All parameters are the same for the continuous blue curves and dashed-dotted green curves  except  $r$ (double for the dashed dotted green curves) resulting in $U=-0.6$ (dashed-dotted green curves) and $U=2.5$ (continuous blue curves). }
\label{fig:plot-V}
\end{figure}

It is instructive to write  \eq{eq:psidot} as $\dot{\psi } =  -\frac{1}{\tau_0}\frac{d}{d \psi}  {V}(\psi) $, in terms of an effective potential ${V}$ with minima corresponding to (meta)stable  phase-locking between the two oscillating units  (\fig{fig:plot-V}).  Motility  requires $\psi\ne0(\pi)$, see~\eq{eq:c-dot-av}. In the absence of chemotactic bias ($\Omega=0$), $V$ presents a single minimum (modulo $2\pi$) if $|U|>1$ (continuous blue curve in  \fig{fig:plot-V}b). A bistable systems with $\psi\ne0(\pi)$ spontaneously emerges from the long-range interactions if $\mathcal{A}>0$ and $ |U| < 1$, with two symmetric stable states corresponding to spontaneous cell polarisation and motion (dashed-dotted green curve in \fig{fig:plot-V}b), separated by an effective energy barrier $\Delta V_0=(1-|U|)^2/2$. Moderate gradients of chemoattractants ($0<\tau_0|\Omega|<1$) introduce a bias that displaces the single minimum ($|U| > 1$) or favours one of the two stable states ($|U| < 1$), (continuous blue and dashed-dotted green curves in \fig{fig:plot-V}d), thus directing motility. 

The  pair of dipolar oscillators considered so far (\fig{sketch}) has  $U> 1 $ and does not exhibit a stable state with broken symmetry. Spontaneous cell polarisation is possible if the force distribution in each unit is itself polarised, the whole cell conserving mirror symmetry. In \fig{fig:plot-V}a, we show an example where each unit consists of two dipoles ($I$ and $II$), separated by a distance $\xi$. To keep the number of parameters minimal, dipole $I$ is chosen to be a scaled version of dipole $II$: $l_I/l_{II}= \mathfrak{R}_I/ \mathfrak{R}_{II}=\kappa$  with $l_{II}=l; \mathfrak{R}_{II}= \mathfrak{R}$. The two dipoles oscillate at the same frequency $\omega$ and with a fixed phase difference; oscillator  $I$ is in opposition of phase with oscillator $II$, which  satisfies \eq{eq:f-dot}. This way, the two units are characterised by a single dynamical phase difference $\psi$  satisfying \eq{eq:psidot}.   As shown in~\cite{SI}, the migration speed for this force distribution is still given by  \eq{eq:c-dot-av} with 
$ 
\Xi  =
 \frac{3  }{32 \pi}   \{ 
 4
  (\kappa^2 - 1 )^2     + 
\frac{ \mathfrak{R}^2 }{l^2} [  \kappa^2   +  1 ]^2 \cos\psi  \}  
$
 and retains the qualitative trend shown in \fig{fig:trend}.  
More interestingly, phase locking at $\psi\ne 0(\pi)$ occurs when $|U|= |\frac{2l^2}{\mathfrak{R}^2}\frac{1-(1-\delta )^3\kappa^2}{1+(1-\delta )^3\kappa^2} | < 1$ (where $\delta :=\xi/r$) and $\mathcal{A} > 0$.
 Correspondingly, $\frac{1}{\tau_0} = \frac{ |A|}{ \mathcal{M} }\frac{  2  a\mathfrak{R}^4 }{ \pi \zeta }  \frac{1+(1-\delta )^3\kappa^2}{r^3 (1-\delta)^3}$.
 
\section{Noisy oscillations result in directed or persistent motions}

Uncorrelated noise,
  associated e.g. to the stochastic nature of actin polymerisation and motor proteins activity,
 influences intracellular synchronisation. Adding a random effective ``force'', $\nu$,  to \eq{eq:psidot}: $\dot{\psi} = \frac{1}{\tau_0}[-\frac{d}{d\psi} {V}(\psi) + \nu]$, results in fluctuations around a stable minimum and occasional jumps between two minima of $V$ leading to  phase slips~\cite{Pikovsky,Gold}. When $|U| >1$, jumps give a phase slip of $2\pi$ and do not change the cell velocity. Velocity fluctuations are controlled by local fluctuation of $\psi$ around the minimum $\psi^*$, estimated to be of order $\langle(\psi-\psi^*)^2\rangle\simeq\Lambda/\partial_\psi^2V(\psi^*)\sim\Lambda/||U|-1|$, where $\Lambda$ is the intensity of the noise, with correlation $\langle \nu(t) \nu (t') \rangle =2 \Lambda \tau_0 \delta(t-t')$. The cell performs a random walk along the $\hh x$ axis in the absence of chemotactic gradients, or a random walk with drift  with such gradients, with a diffusion coefficient of order $D\simeq\langle v_c^2\rangle\tau_0\sim(\partial_\psi v_c(\psi^*))^2\tau_0\Lambda/(|U|-1)$ (continuous blues curves in \fig{fig:plot-V}c,e).
When  $|U|  < 1$,   the two stables states $\psi_\pm$  (with $\sin \psi_- = -\sin \psi_+$ ) have non-vanishing velocities $v_c(\psi_-)=-v_c(\psi_+)$ (\eq{eq:c-dot-av}). 
The cell's trajectory is thus a succession of persistent motions separated by random changes of direction, resembling the run and tumble motion of bacteria \cite{berg:1972}. The  persistence time of directed motion is the first passage time between two consecutive jumps: $\mathcal{T}_{\pm} \propto  \tau_0      e^{\Delta V_{\pm}/\Lambda} $~\cite{Hanngi}.  
Typical trajectories without and with chemotactic gradients, obtained from numerical simulations (see~\cite{SI} for details), are shown in \fig{fig:plot-V}c,e (dashed-dotted green curves). 

\section{Discussion}

Our model  captures within a single framework two 
 distinct chemotactic strategies, namely the case where the gradient polarises an otherwise quiescent cell  ($|U|>1$) and  the case where the gradient biases a preexisting polarised state showing transient directed motion ($|U|<1$). Interestingly, in the latter case the directional bias introduced by chemotactic gradients is mostly due to a bias in the ``run'' time $\mathcal{T}_\pm$, due to the asymmetry of the potential barrier ($\Delta V_->\Delta V_+$ in \fig{fig:plot-V}e), and only slightly to directional bias in the cell velocities ($|v_c(\psi_-)|>|v_c(\psi_+)|$). This chemotactic strategy, reminiscent of the one of bacteria \cite{berg:1972}, has to some extent been observed for {\em Dictyostelium} \cite{Insall,amselem:2012}.

The value of  $|U|$ is critical to spontaneous cell polarisation, the persistence of cell motion, and the cell chemotactic response. Factors that affect the value of $U$ will thus qualitatively affect the cell's trajectories. For a given force distribution in each unit at the cell edges, increasing the cell size $r$ monotonously decreases $U$. The condition $|U|<1$ corresponds to a range of cell size, and we thus predict that very small or very large cells (compared to the size of one force unit) should not  show spontaneous polarisation. Another remarkable result is that in the fixed force regime (when $\mathfrak{R}\propto 1/\zeta$), the parameter $U\sim\zeta^2$ increases with increasing substrate viscosity. We further predict that not only is the instantaneous cell velocity affected by the substrate mechanics as depicted in \fig{fig:trend}, but also the {\em persistence} of cell motion, and the ability for spontaneous cell polarisation in the absence of external cues, could be strongly impaired under high substrate mechanical resistance.
\acknowledgements
The work was supported by the Human Frontier Science Program under the grant RGP0058/2011. We  thank 
B. Ladoux, M. Sixt and T.B. Liverpool (M.L.) for stimulating discussions.


\appendix

\section{Flow   at the cell-substrate interface}
 \label{sec:flow-interface}
 
\subsection{Two coupled fluid layers} 
\label{sec:two-fluid-layers}

 In this Appendix we outline  
   the  procedure to obtain the 
   analytical solution of~\eq{eq:coupled-fluids}  considered
  in the main text.
  Related problems were studied previously  also by other authors, see e.g. ref~\cite{Lenz,Lubensky}.

 We consider two planar fluid interfaces, 
    at position $z=0$ 
and at position $z= \mathfrak{h}$
where   $\hh z$  is a unit vector  indicating
     the direction normal to the interface.
To proceed it is useful to decompose
the velocity appearing in~\eq{eq:coupled-fluids} 
 $\m v^\varsigma(x, y, z) = \m v^\varsigma_\parallel(x, y, z) + 
v^\varsigma_z(x, y, z)  \hh z$ in terms of
the in-plane components  $\m v^\varsigma_\parallel(x, y, z)  $ and orthogonal
component $v^\varsigma_z(x, y, z) $  
where the label $\varsigma = c, s$ indicates either cell or the substrate.
A similar decomposition can be done for the gradient:
$\bm{\nabla} = \bm{\nabla}_{\parallel} +\hh z \frac{\partial }{\partial z} $
where 
$\bm{\nabla}_{\parallel}  = (\frac{\partial }{\partial x} ; \frac{\partial }{\partial y})  $
has components only  in the $(x, y)$ plane.
Finally the Laplacian
becomes
$ \nabla^2 =  \nabla^2_\parallel + \frac{\partial^2 }{\partial z^2} $.

The boundary conditions at the interfaces are as follows:
no motions along z-axis,
$\m v^{(c)} \cdot  \hh z |_{z=\mathfrak{h}} = 0$, 
and continuity of the tangential component of the velocity at the interface; 
no-stress condition at the cell-upper solvent interface (obtained neglecting solvent viscosity)
$ 
 \eta_c [ \bm{\nabla}_\parallel v^{(c)}_z + \frac{d}{d z} \m v^{(c)}_\parallel ]_{z=\mathfrak{h}}
 =  0.
 $
At the cell-substrate interface,  $z=0$,  
the conditions are
no z-motions of the interface,
$\m v^{(c)} \cdot  \hh z |_{z=0} = 0$
and 
$\m v^{(s)} \cdot  \hh z |_{z=0} = 0$;
continuity of the 
tangential  component, 
$ \m v^{(c)}_\parallel |_{z= 0}= \m v^{(s)}_\parallel |_{z= 0}$; 
stress balance, 
$ 
\eta_c [ \bm{\nabla}_\parallel v^{(c)}_z + \frac{d}{d z} \m v^{(c)}_\parallel ]_{z=0}
-
 \eta_s [ \bm{\nabla}_\parallel v^{(s)}_z + \frac{d}{d z} \m v^{(s)}_\parallel ]_{z=0}
=
-\bm{F}( \m r, z= 0).
 $
 As explained in the main text, 
 it  suffices to focus on a single interfacial point force
 $\bm{F}( \m r, z = 0) = \m f \delta(\m r)$.

 We search a solution using Fourier transforms.
By exploiting  the geometry of the system, we  
 decompose the Fourier vector $\m k$ as $\m k = \m q + p \hh z$
where $\m q $ lies in the plane $(x, y)$ orthogonal to direction $\hh z$.
Thanks to this decomposition, 
a generic vector  field $\m E(x,y,z)$ in three-dimensions
 in presence of the
 planar interface at $z=0$ 
can  be written as 
\begin{align}
& \m E(x,y,z) = \int \frac{ d^3\m k}{(2\pi)^3} e^{-i \m k \cdot \m x}   \tilde{\tilde{\m E}}(\m k)
=
\int \frac{ d^2\m q }{(2\pi)^2} e^{-i \m q \cdot \m r}     {\tilde{\m E}}(\m q,z).
\end{align}

  We proceed by taking the Fourier transform
of the  velocity components.
The
condition of  no motion in the z-direction at the interfaces 
means $ \tilde{ v}^{(\varsigma)}_z(\m q, 0) = 0  $ and  $ \tilde{ v}^{(c)}_z(\m q, \mathfrak{h} ) = 0  $
 so  the various stress conditions  only involve  terms 
  $\tilde{\m v}^{(\varsigma)}_\parallel(\m q, z)$ for $\varsigma= c, s$.  

We now make an ansatz concerning the   
z-dependence of the Fourier components. 
For the flow in the substrate $z \le 0$
 we pose 
$\tilde{\m v}^{(s)}_\parallel (\m q, z) :=  \m S(\m q) e^{q z}.$
For the intracellular flow,
 in the  region $0< z < \mathfrak{h}$, we pose 
$
\tilde{\m v}^{(c)}_\parallel (\m q, z) =  \m C_+ (\m q) e^{-q z}
+ \m C_- (\m q) e^{q z}.
 $
 The three
 terms $\m C_{\pm}, \m S $
 can be determined using the  boundary conditions for the stress and  velocity given before.
 
 As we have three unknowns,
it suffices to consider three of the four equations for the tangential velocities discussed above.
We choose 
the continuity of the tangential velocity at at $z=0$,
  $\m C_+ + \m C_- = \m S $
and the two equations for the interfacial stresses 
 at $z=\mathfrak{h} $:
 $
0
=
 \eta_c [-\m C_+ e^{-q  \mathfrak{h} }
+ \m C_- e^{q  \mathfrak{h}  } ] ;
$
at $z=0$:
 $
 q \eta_c [- \m C_+
+ \m C_-  ]
-
q \eta_s \m S
 = - \m f .
$
From the second  equation we  obtain
$\m C_+ =    e^{2 q \mathfrak{h} } \m C_- $.
Using this relation in the first equation
we get
$\m S =\m C_- (1 +e^{2 q \mathfrak{h} } )  $.
Inserting these relations  in the third equation, 
we determine $\m C_-, \m C_+$
and finally
the flow at the cell-substrate interface $z=0$
$$  \tilde{\m v}^{(s)}_\parallel(\m q, 0)
  \equiv  \m S \equiv \m C_+ + \m C_- = 
\frac{ \m f}{ q \{
  \eta_c [\tanh (\mathfrak{h}  q) ]  
+
 \eta_s   
 \}  } .
 $$
 
 The inverse Fourier transform of $\m S(\m q)$
  has two analytically tractable limits :  
   $\mathfrak{h} \to  0$, 
corresponding to a thin-film, 
 and 
 $\mathfrak{h} \to  \infty $,
 representing two semi-infinite coupled fluids. 
 We  focus on the latter 
 where $[\tanh (\mathfrak{h}  q) ]  = 1 $
and 
\begin{align}
&\m v^{(c)}_\parallel(\m r, 0)  \equiv \m v^{(s)}_\parallel(\m r, 0) = \frac{\m f}{(2 \pi)^2(\eta_s+\eta_c) }\int d^2 \m q \frac{1}{q} e^{-i \m q \cdot \m r}
\nonumber \\
&
=
\frac{\m f}{2 \pi (\eta_s+\eta_c)} 
\int^\infty_0 d q J_0(q r) 
= \frac{\m f}{2 \pi (\eta_s+\eta_c) r}
\label{eq:flow}
\end{align}
Here  we used 
$\int d^2 \m q \frac{1}{q} e^{-i \m q \cdot \m r}   =
\int^\infty_0 d q q \int^{2\pi}_0 d \theta \frac{1}{q} e^{-i  q   r \cos \theta} 
=2 \pi \int^\infty_0 d q J_0(q r) 
  $ where $J_0$ indicates the modified Bessel function of the first kind of zeroth order~\cite{Abra}.
\eq{eq:flow} is the expression 
reported in the main text and
represents   
the disturbance in the flow field
at the interface of  two semi-infinite fluids,
  generated  at position $\m r$ by a point force
 placed  at the origin, and $r := \Vert \m r \Vert$.
We can  write this relation
 as $\m v(\m r) = H(\m r) \m f$
  where
$H(\m r) $ is the 
Greens's function describing the flow at the interface.
 Thanks to the linearity of the equations 
describing the fluids,
the effect of $N$ point forces is obtained
 by superposing the  single  effects.
 The velocity disturbance generated at the interface
  at position $\m x$   
   due to    $N$ disks, each one centred  at position $\m x_n$
  and subjected to
 forces $\m f_n$ with $n = 1, \ldots N$,
is  given by
$\m v(\m x) 
= \sum^N_{n=1}  \frac{1}{2 \pi (\eta_c +\eta_s) \Vert \m x- \m x_n \Vert}  \m f_n
.$

\subsection{Narrow gap  between two  fluid layers}
\label{sec:narrow-gap}
\begin{figure}[h!]
  \centering
  \includegraphics[width=0.3\textwidth]{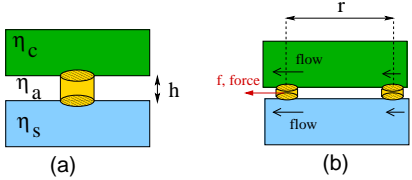}
  \caption{
  (Color online)
  (a) scheme of a viscous cell of viscosity $\eta_c$, lying of a viscous substrate of viscosity $\eta_s$
separated by a small gap $h$ of viscosity $\eta_a$. The disks represent
 force-generating elements which are in contact with
both the cell and the substrate. (b) When subjected to a force, each disk drags its surrounding molecules and generates a local flow. This flow propagates and affects also other disks far apart.
   }
  \label{fig:3fluids}
  \end{figure}

In this section we study the case where the cell
interacts with the substrate only through discrete sites (disks)
 corresponding to the force-generating elements
while the remaining part of the interface is allowed to slip  ($\m v^{(c)} \neq \m v^{(s)}$)  without friction.
To this end we consider a narrow gap,
a quasi-2D film of height $h$ and small viscosity $\eta_a$,
 separating 
the two fluids (see~\fig{fig:3fluids}). 
For simplicity we shall treat the film as a 2D fluid~\cite{Lubensky,Levine,LLEPL2010}.
The coupled Stokes equations
 now read 
\begin{align}
& \eta_c \nabla^2 \m v^{(c)} -\bm{\nabla} p^{(c)} = 0; 
\hspace{0.2cm}
\textrm{with}
\hspace{0.2cm}
 \bm{\nabla} \cdot \m v^{(c)} = 0;
 \hspace{0.2cm}  z > 0 \nonumber \\
& \eta_a \nabla^2_\parallel \m v^{(a)} -\bm{\nabla}_\parallel p^{(a)} 
+ \eta_c \partial_z \m v^{(c)} - \eta_s \partial_z \m v^{(s)}
= -\m f \m \delta( \m x)  ; 
\nonumber \\
&
\hspace{0.2cm}
\textrm{with}
\hspace{0.2cm}
 \bm{\nabla}_\parallel \cdot \m v^{(a)} = 0;
 \hspace{0.2cm}  z = 0 \nonumber \\
& \eta_s \nabla^2 \m v^{(s)} -\bm{\nabla} p^{(s)} =  0 ;
\hspace{0.2cm}
\textrm{with} 
\hspace{0.2cm}
\bm{\nabla} \cdot \m v^{(s)} =0 ;
\hspace{0.2cm}
 z <0 \label{eq:stokes22}
\end{align}
where  
 $\m v^{(c)}$, $p^{(c)}$ indicate
velocity and pressure 
  for the intracellular flow; 
 and  $\m v^{(s)}$, $p^{(s)}$
velocity and pressure 
  for the substrate flow as before. $\m v^{(a)}$, $p^{(a)}$ are the 2D
velocity and pressure 
  in the film, so  $\eta_a$ is a two-dimensional viscosity
  and the ratio $\eta_a/(\eta_c+\eta_s) $ has the dimension of a  length~\cite{Saffman,Lubensky,Levine,LLEPL2010}.
  The terms $\eta_c \partial_z \m v^{(c)}$ and $- \eta_s \partial_z \m v^{(s)}$
  describe the stress exerted by the cell and the substrate on the film. Equal and opposite stresses are exerted by the film on the upper and lower bulk fluids, which enter as boundary conditions for the first and the last line of~\eq{eq:stokes22}.

We
  impose the continuity of the velocities
at the location of the disks, 
$\m v^{(c)}|_{\textrm{disk}} = \m v^{(s)}|_{\textrm{disk}} = \m v^{(a)}|_{\textrm{disk}}  $.
This way, using the ansatz
$ \tilde{\m v}^{(s)}(\m q) = \m S(\m q) e^{z q} $; $ \tilde{\m v}^{(c)}(\m q) = \m C(\m q) e^{-z q} $ ;
$ \tilde{\m v}^{(a)}(\m q) = \m A(\m q)  $
we get  
the equation in Fourier space for the film. 
Taking the inverse Fourier transform,
$ \m v^{(a)}(\m r) = \int \frac{d^2\m q}{(2\pi)^2} e^{-i\m q \cdot \m r} \frac{ [\mathbb{I} -\hh q \otimes \hh q]}{[ q^2 \eta_a  
 + q (\eta_c + \eta_s) ] }   \cdot \m f$. 
 The integral can be calculated analytically,
 the result is expressed in terms of special functions.
 The limit of negligible film viscosity 
  can be obtained directly
 by setting  $\eta_a =0$ in the above expression.
 The flow  $\m v^{(a)} =v^{(a)}\hh x$  resulting from a force in the same direction $\m f = f \hh x$ is again of the form
  $ v^{(a)}(r) \sim  \frac{1}{(\eta_s +\eta_c) r} f. $

\subsection{Equations describing  elastic cells  lying on  viscous substrates}
\label{sec:elastic-cell}
Here we  generalise our description to the case of  an elastic cell characterised by bulk elasticity $\mu_c$. 
The disks are arranged with
 the same geometry considered in~\fig{sketch}, they
oscillate around their equilibrium positions of~\fig{sketch} 
but are now   permanently bound  to the cell elastic body.  
The dynamics at the interface ($z=0$) is derived as  in~\sec{sec:two-fluid-layers}. 
 In absence of motions along $z$-axis,
the force balance is described by
$\mu_c \partial_z \m u^{(c)}_{\parallel} -\eta_s \partial_z \m v^{(s)}_\parallel = - \m F(\m r, z, t)$
where $\m u^{(c)} $ represents the deformation field in the elastic medium. 
The decomposition of $\m u^{(c)} $ into  components that are 
parallel, $\m u^{(c)}_\parallel $, and orthogonal,  $ u^{(c)}_z $, to the interface is defined as
 in~\sec{sec:two-fluid-layers}. 
 The velocities in the
upper medium (cell) and 
 in the lower medium (substrate) coincide 
 at the locations of the disks
\begin{align}
 \partial_t \m u^{(c)}_{\parallel}|_{\textrm{disk}} = 
\m v^{(s)}_\parallel |_{\textrm{disk}} .
\label{eq:vel-disk-location}
\end{align} 
Once the force-distribution is known, these equations determine
 both deformation in the cell and the flow  generated in the substrate at the location of the disks.

Once again to proceed we define Fourier transforms, now  
with  respect to both space and time, 
$g(\m x, t) = \int \frac{d^2 \m q d\Omega}{(2 \pi)^4} e^{-i \m q \cdot \m r} e^{-i \Omega t} \tilde{g}(\m q, \Omega) $
for any function $g$. From~\eq{eq:vel-disk-location},  
  $-i\Omega \tilde{\m u}^{(c)}_\parallel = \tilde{\m v}^{(s)}_\parallel $.
 To obtain the $z$-dependence of the solutions
  we proceed as in~\sec{sec:two-fluid-layers},  posing 
$\tilde{\m v}^{(s)}_\parallel(\m q, z, \Omega) = \m S(\m q, \Omega) e^{q z}  $ and  $\tilde{\m u}^{(c)}_\parallel(\m q, z, \Omega) = \m C(\m q, \Omega) e^{-q z}  $. Therefore at the disk locations
$  -i \Omega \m C(\m q, \Omega) = \m S(\m q, \Omega) $, resulting  in  
 $\m S(\m q, \Omega) = \tilde{\m F}(\m q, \Omega)/ [ q( \eta_s   + i \frac{\mu_c}{ \Omega} ) ] .$

As we are interested in the oscillatory behaviour of the cell,  
we consider a point-like force which oscillates with frequency $\omega$, $\m F (\m x, t) = \m f_0 \delta(\m x) \sin(\omega t)$. 
As a result $\tilde{\m F}(\m q, \Omega) = \m f_0 \frac{2 \pi}{2 i}[ \delta(\Omega + \omega) -  \delta(\Omega -\omega)] $.
Using this we can take the inverse transform 
and obtain the velocity field at the interface
at the position of a disk as (for simplicity we remove superscripts)
$$ \m v_\parallel(\m x, t) = \int \frac{d^2 \m q d\Omega }{(2 \pi)^2} \frac{e^{-i \m q \cdot \m r} e^{-i \Omega t}}{q [\eta_s + i \frac{\mu_c}{\Omega}] }  \frac{\m f_0 }{2 i}[ \delta(\Omega + \omega) -  \delta(\Omega -\omega)]. $$
Performing the integrals, we can recast the resulting expression using Green's function as
\begin{align}
& \m v(\m x, t) = \bm{H}^{(e)}(r)\cdot  \m F + \partial_t[\bm{G}^{(e)}(r) \cdot \m F]; \label{eq:v-elast-visc}
   \\
 & {H}^{(e)}_{ij}(r) = \frac{\eta_s}{\eta^2_s + \frac{\mu^2_c}{\omega^2}} \frac{1}{2 \pi r} \delta_{ij} ;
\hspace{0.3cm}
 {G}^{(e)}_{ij}(r) = \frac{\frac{\mu_c}{\omega} }{\eta^2_s + \frac{\mu^2_c}{\omega^2}} \frac{1}{2 \pi r} \delta_{ij}
\nonumber 
\end{align}
where $\delta_{ij}$ is the Kronecker symbol.
We note that letting $\mu_c \to 0$ one recovers the expression for the semi-infinite viscous substrate, and similarly
letting $\eta_s \to 0$ one obtains the equivalent expression for the elastic medium.

\section{Dynamics of the four disks system}
We focus on the
  system made up of  two oscillating units, labeled with $\alpha = A, B$, 
 each unit consisting of a single dipole discussed in the text, see~\fig{sketch}. 
Although in the main text we restricted to  a one-dimensional force distribution,
 in the following section we derive the equations in  vectorial form valid for generic
  force distributions.
  
\subsection{Derivation of the equations describing the crawler's dynamics}
\label{sec:crawl-eq}

The dynamics at the interface
 of two 
  fluids with negligible inertia  
 and viscosities
$\eta_c$ and $\eta_s$ 
is instructive to
highlight similarities and differences between the  locomotion of a soft object
and the  swimming in  bulk viscous fluids at low Reynolds number~\cite{Purcell}.

We begin with noting that 
 the motion of a  disk 
 lying 
at the interface of the two fluids   can be very complicated : 
 a  disk straddling  the interface can 
  cause deformations or instabilities of the interface
 which in turn affect the disk motion. 
 A simple approximation consists in \emph{neglecting} this aspect of the interface dynamics.
 Such an approximation is  valid if the surface tension is constant and sufficiently 
 large  to prevent any interface deformation. 
 This simplified version of the problem  has been studied and solved by Ranger~\citep{Ranger}. 
 The   analogous problem for the sphere was discussed more recently by Pozrikidis~\cite{Poz07}.

A soft cell migrating over a substrate has  peculiar mechanical features,  which we now explain.
The cell-substrate interface divides the system into two, 
 individuating
 intracellular 
 and extracellular forces and flows.
Intracellular  forces originate from the active behaviour of the cell and the cytoskeleton.
They are responsible for cell oscillations
 but  do not contribute directly to the cell migration,
 due to the confining effect  of the cell boundary.
In our model, only  substrate-mediated interactions are responsible for net 
   translational motion of the cell relative to the substate, which occurs thanks
    to the stress transmitted via the substrate.
However,  both the dissipation  associated to the intracellular forces and that
 associated to the extracellular forces must be included as local contributions. 
This is at the origin of the biphasic behaviour as a function of the substrate viscosity in our model.

To see this, we  consider the four-disks system
of ~\fig{sketch}. 
We write the force balance for the disks as
\begin{equation}
 \m F^\alpha_n = \zeta (\dot{\m x}^\alpha_n -\m v(\m x^\alpha_n)) +\bm{\lambda}^\alpha_n 
 \label{eq:f-bal-constrN}
\end{equation}
where
the new terms $\bm{\lambda}^\alpha_n $ 
are Lagrange multipliers
that  prevent net motion of the centre of the collection of  disks with respect to
 the cell frame, representing  e.g. the interaction with 
  the cell boundary which is not explicitly included in our model. 
 As in the main text, $\zeta = \zeta_s +\zeta_c$
 is the total drag coefficient ;
 $\dot{\m x}^\alpha_n $ is the velocity of particle $n$ ;
 $\m v(\m x^\alpha_n)$ is 
  the flow at particle $(\alpha,  n)$ due to all the remaining force centres.
  
We identify
 the non-local, intracellular-mediated,
  force acting on particle $(\alpha,  n)$ :
   $\bm{\mathcal{C}}(\m x^\alpha_n) := \zeta_c \m v(\m x^\alpha_n)$.
    This force is
    felt
    by the disk $(\alpha,  n)$ at position
  $\m x^\alpha_n$, due to all the remaining particles and represents 
      a non-local effect
       mediated by the intracellular environment.
Similarly,
we identify
 the non-local,  substrate-mediated,
  force acting on particle $(\alpha,  n)$ as
   $\bm{\mathcal{S}}(\m x^\alpha_n) :=\zeta_s \m v(\m x^\alpha_n )$. 
  The meaning is analogous:
$\bm{\mathcal{S}}(\m x^\alpha_n)  $ is the  force  felt by
 particle $(\alpha,  n)$ (at position  $\m x^\alpha_n$)
  due to all the remaining particles representing
      a non-local effect  mediated by the substrate (extracellular environment).

 We  write~\eq{eq:f-bal-constrN} inserting these
  definitions of
 non-local intracellular and extra cellular forces and rearranging terms,
  $  \zeta \dot{\m x}^\alpha_n 
   =\m F^\alpha_n  
  +  \bm{\mathcal{C}}(\m x^\alpha_n) 
  + \bm{\mathcal{S}}(\m x^\alpha_n)
  -\bm{\lambda}^\alpha_n. $
The Lagrange multipliers $\bm{\lambda}^\alpha_n$ are determined
 by 
requiring  
:
i)  
 no net relative motion between
   the centre of the system of oscillating units
  and the cell boundary ;
 ii) 
   that the oscillatory dynamics of each dipole $d_\alpha$ remains unaffected.

Condition i) is implemented by taking
the sum over all the particles in the previous equation.
By definition the system is force-free, so the sum of all the active forces driving the particles vanishes,
 $ \sum^2_{n = 1} \sum_{\alpha = A, B}  \m F^\alpha_n   = 0 $. Again by definition,
  $  \sum^2_{n = 1} \sum_{\alpha = A, B}    \dot{\m x}^\alpha_n = 4  \dot{\m c} $.
 Condition i)  reads 
$   \sum^2_{n = 1} \sum_{\alpha = A, B}   \bm{\mathcal{C}}(\m x^\alpha_n)
  = \sum^2_{n = 1} \sum_{\alpha = A, B}     \bm{\lambda}^\alpha_n $. Together 
  these result in
\begin{equation}
\dot{\m c} = \frac{1}{4} \frac{ \zeta_s }{(\zeta_s +\zeta_c) }
 \sum_{\alpha = A, B} \sum_{n  = 1,2} 
 \m v(\m x^\alpha_n)
\label{eq:cdot-cell-4disks}
\end{equation}   
In addition, condition $ii)$  implies $\bm{\lambda}^\alpha_1 = \bm{\lambda}^\alpha_2 $ for $\alpha = A, B$ so
each oscillating unit satisfies 
\begin{equation}   
  \zeta  \dot{\m d}_\alpha
 =    \m F^\alpha_1 - \m F^\alpha_2
  + \zeta [  
  \m v(\m x^\alpha_1) 
-  
 \m v(\m x^\alpha_2) ].
  \label{eq:ddot-cell-4disks}
\end{equation}   
 Analogous expressions can be obtained for the eight disks system, see~\cite{SI}.

 \subsection{Elastic cells on viscous substrates}
 \label{sec:elastic-cell2}
We return to the case of  disks bound to an elastic cell body.
 The equation describing the dynamics of one isolated disk, centred at position $\m x^{(eq)}$, 
 at the interface 
between  elastic cell and  viscous substrate  
 is given by 
 $-\zeta_s \dot{\m x} - \xi_c [\m x-\m x^{(eq)}] + \m f = 0$. Here $\zeta_s$ 
 describes the substrate viscous drag as before and
 $\xi_c$ the coefficient relating  deformation and force  applied to the centre of 
 a  disk~\cite{Lin} lying at  the interface (for a disk of radius $a$, 
  $\xi_c = \frac{16}{3} a \mu_c $).
  This expression can be readily generalised
 to the case of many disks, labeled as before with indices $\alpha$ and $n$, 
 as $-\zeta_s [\dot{\m x}^\alpha_n -\m v(\m x^\alpha_n)] - \xi_c  [ \m x^\alpha_n- \m x^{(eq), \alpha}_n -\m u(\m x^\alpha_n) ] + \m F^\alpha_n = 0$.
 $\m v(\m x^\alpha_n)$ and $\m u(\m x^\alpha_n)$ represent the velocity field in the substrate
 and the deformation field in the cell
  at the location of disk $\alpha, n$. 
   From these equations the  dynamics of each oscillating units $\alpha = A, B$ is obtained as
\begin{equation}
\dot{\m d}_\alpha = \frac{[\m F^\alpha_1 -\m F^\alpha_2]}{\zeta_s} 
-\frac{\xi_c }{\zeta_s} \m d_\alpha 
+ \m v(\m x^\alpha_1) - \m v(\m x^\alpha_2) + \frac{\xi_c }{\zeta_s}[\m u(\m x^\alpha_1)- \m u(\m x^\alpha_2)]
\label{eq:ddot-elast-cell}
\end{equation}
  
Since the disks are rigidly connected to the elastic body, 
the terms $\bm{\lambda}^\alpha_n$ 
used above to enforce the average disk locations with respect to the cell boundaries 
are not needed.
   The migration speed is obtained  by studying the motion of 
  tracers lying in the substrate, below the disks, at  $z \to 0^{-} $. 
  We need four tracers for the four disks system and eight tracers for the eight disks. 
The  tracers are  convected by the flow underneath the disks so the tracer dynamics is
 described by 
$ - \zeta_s [\dot{\m x}^\alpha_n - \m v(\m x^\alpha_n)] = 0$. 
Thus
the centre of the collection of tracers moves according to
\begin{equation}
\dot{\m c} = \frac{1}{4} \sum_{\alpha = A, B} \sum^{2}_{n=1} \m v(\m x^\alpha_n) 
\label{eq:cdot-elast-cell}
\end{equation}
 where 
 $\m v(\m x^\alpha_n)$ is given by~\eq{eq:v-elast-visc}.
This expression describes  the propulsive speed of the cell.

\subsection{Force-dipole and quadrupole}
From now on, we restrict ourselves to the one-dimensional force distribution discussed in the main text.
 Just like in the main text, the disks are centred at positions $x^{\alpha}_n$ and subjected to  forces 
 $F^{\alpha}_n$ which  satisfy $f^{\alpha} := F^{\alpha}_1
  = -F^{\alpha}_2$, for $\alpha=A,B$.
 The  coordinates can be written as 
\begin{align}
& x^A_{n} = c -\frac{r}{2} + (-)^{n+1}  \frac{ L_A}{2};
\qquad
 x^B_{n } =c+ \frac{r}{2} + (-)^{n+1} \frac{L_B}{2} \label{eq:coord} 
\end{align}
where $r = c_B - c_A$ is the separation between the  centres of the two oscillating units,  see~\fig{sketch}, and $(-)^n = (-1)^n $.
The quantities $c$ and $L_{\alpha}$
are the crawler's centre and the (time-dependent) length of dipole $\alpha$.
With no  loss of generality we pose  $L_{\alpha } = l_\alpha + d_{\alpha}$ where $l_\alpha $ is a constant and $d_\alpha$ is time-varying as in~\eq{eq:param-d-f}.

 The force-multipoles are
 defined as moments of the force distribution. 
To illustrate this we consider~\eq{eq:coord}. The moment of  order $k$  is
$ \mathcal{M}^{(k)} : =\sum_{\alpha = A, B } \sum_{n=1,2} F^\alpha_{n } (x_{n}^\alpha -c)^k  $.
We pose   
  $\mathcal{D} := \mathcal{M}^{(1)}$, for the dipole,  and 
$\mathcal{Q} := \mathcal{M}^{(2)} $, for the quadrupole,
 and find that
 they are  given by the matrix relation
\begin{equation}
 \begin{pmatrix}
  \mathcal{D}\\
  \mathcal{Q}
 \end{pmatrix}
 = \begin{pmatrix}
 L_{A} &  L_{B}  \\
 - r L_{A} &  r L_{B}
 \end{pmatrix}
 \cdot
 \begin{pmatrix}
 f_{A} \\
  f_{ B} 
 \end{pmatrix}
\end{equation}
 showing that the total dipole $\mathcal{D}$ of the system 
is the sum of the two individual dipoles.  
In turn, the two  forces $f_A$ and $f_B$ 
 can be expressed as functions of dipole and quadrupole moments as
\begin{equation}
 \hspace{0.4cm} 
\begin{pmatrix}
 f_{A} \\
  f_{B} 
 \end{pmatrix}
 =\frac{1}{2} \begin{pmatrix}
 \frac{1}{ L_{A} } &  -\frac{1}{ r L_{A} }  \\
  &  \\
  \frac{1}{ L_{B} } &  \frac{1}{ r L_{B} }
 \end{pmatrix}
 \cdot
 \begin{pmatrix}
  \mathcal{D} \\
  \mathcal{Q} 
  \label{eq:fA-fB-D-Q}
 \end{pmatrix}.
\end{equation}
This relation implies that   higher moments $\mathcal{M}^{(n)}$,
 with $n > 2$, can be expressed as combinations of  the two moments $\mathcal{D}$ and 
 $\mathcal{Q}$.

\subsection{Migration speed}
\label{sec:mig-speed}
To begin with, we note that the
   flow generated 
  at  position $x^\alpha_n$
(the centre  coordinate of the generic disk $(n, \alpha)$)  
  by all the remaining disks can be written as 
\begin{equation}
 v(x^\alpha_n) = \sum_{m \neq n}
 H_{n\alpha m \alpha}
 F^\alpha_m
+ \sum_{\beta \neq \alpha} \sum_{m =1,2} 
 H_{n\alpha m \beta }
 F^\beta_m
\label{eq:vH}
\end{equation} 
where we have introduced
 $H_{n\alpha m \beta } := H(x_{n}^\alpha -x_{m}^\beta ) \equiv \frac{1}{2\pi (\eta_s +\eta_c) | x^\alpha_n  - x^\beta_m |} $
  as obtained in \sec{sec:two-fluid-layers}.
 
\subsubsection{Instantaneous migration speed}

From the force balance, using~\eq{eq:cdot-cell-4disks} 
and the expression in~\eq{eq:vH}
we obtain the instantaneous migration speed
 for the system  depicted in~\fig{sketch} 
as
\begin{align}
&  \dot{c} 
= \frac{\zeta_{s} }{\zeta} \frac{1}{4} [ (H_{1A1B} -H_{1A2B} +H_{2A1B} -H_{2A2B} ) f_B  
\nonumber \\
&
+
(H_{1B1A} -H_{1B2A} +H_{2B1A} -H_{2A2B} ) f_A 
].
\end{align}
\subsubsection{Approximate expression for instantaneous migration speed}
An approximate analytical  expression for the migration speed
 is obtained by performing the analogue of a multipole expansion,
valid for $2 r \gg  | L_A + L_B |$, up to third order in the separation $r$ 
\begin{align}
&  \dot{c}
\approx \frac{\zeta_{s}}{\zeta} \frac{1}{4}
 \Big\{2 H'  [ L_B f_B -  L_A f_A]
 \label{eq:c-dot-inst} \\
&
   + \frac{1}{12}  H''' \big[ ( L^3_B + 3 L^2_A L_B )    f_B  
-
    ( L^3_A + 3 L_A L^2_B  ) f_A   \big]  
\Big\} \nonumber 
\end{align}
where we defined $H' := \frac{d}{d r} H(r) $ and $H''' := \frac{d^3}{d r^3} H(r) $.

\subsubsection{Instantaneous migration speed 
and dipole/quadrupole moments}
We now briefly
 discuss how the migration speed is related to
the dipole and quadrupole term of the force-distribution.
Inserting the expression of $f_A $ and $f_B$ as functions of
$\mathcal{D} $
and
$\mathcal{Q} $  obtained from the second relation  of~\eq{eq:fA-fB-D-Q}
we obtain  
\begin{align}
&  \dot{c} 
\approx \frac{\zeta_{sub}}{2 \zeta} 
 \Big\{   H'   \frac{\mathcal{Q}}{r } 
   + \frac{H'''}{24}   \big[  
(   L^2_A - L^2_B)
     \mathcal{D} 
   +2  ( L^2_A +  L^2_B  ) 
\frac{\mathcal{Q}}{r }     
       \big]  
\Big\}.
\label{eq:c-dot-inst2}
\end{align}
\eq{eq:c-dot-inst2}
shows that,  
to leading order in the separation $r$, 
the sign of the instantaneous migration speed
 is determined by the quadrupole, consistently with
what reported in~\cite{Tanimoto14}. 
Since here 
 the terms $f_\alpha$ and  $d_\alpha$ with $\alpha = A, B$ are oscillating quantities with
 zero mean, the same holds for  dipole and quadrupole $ \mathcal{D}, \mathcal{Q} $.
 As a consequence, also the first term
  of~\eq{eq:c-dot-inst2} 
  oscillates   
 with zero mean
 and does not contribute to the net migration speed.
A net contribution
 comes instead from  the remaining terms
 in~\eq{eq:c-dot-inst2}
 which depend again  on 
 dipole or quadrupole but are more involved.   
 Ref.~\cite{Tanimoto14} does not report how the average migration speed depends on the dipole or quadrupole so a direct comparison with experiments is not yet available in this case.

\subsubsection{Approximated expression for the average migration speed}
  \label{sec:viscous-cell}
To obtain the average migration speed 
we insert the
  parametrisation of~\eq{eq:param-d-f} in~\eq{eq:c-dot-inst}
   and take the average over the period $T := \frac{2\pi}{\omega}$.
In doing so, for example, 
we find that the term
$L_A f_A- L_B f_B $
contains a combinations of $e^{i\omega t}$ and $e^{-i\omega t}$
and therefore
 vanishes when we average over the  period  $T$ as discussed above.
For the same reason, the term $L^3_A f_A- L^3_B f_B $ does not contribute.
Instead, the remaining term
gives a finite contribution to the average as
$  L_A L_B ( L_A f_B  -  L_B f_A)  \sim [2 l^2 
 + d_A d_B ]( d_A  f_B  -  d_B  f_A)   $
where $\sim $ neglects terms that  average to zero.
In particular, 
we find
 $d_A f_B- d_B f_A 
 \sim 
-\frac{1}{2} \zeta \omega R_A R_B\sin \psi
$ 
where 
$\psi := \phi_B -\phi_A$. 
Similarly
$ d_A d_B [d_A f_B- d_B f_A ]
 \sim
   -\frac{\zeta \omega}{8}  R^2_A R^2_B \sin 2 \psi.
$
Hence,
 to leading order in the multipole expansion
  the average migration
speed in a period $T$ 
is   the 
 expression
 reported in the main text,~\eq{eq:c-dot-av},
 where 
\begin{align}
  \Xi  & := \frac{3}{32 \pi} [ 4  + \frac{R_A R_B}{l^2} \cos \psi ]   
\label{Xi-4beads} \\  
&
=
  \frac{3}{32 \pi} [ 4  + \frac{4 g_A g_B}{l^2 \omega^2 \zeta^2}   \cos \psi ]. 
\nonumber 
\end{align}
 The last equality holds
  by virtue of the relation among deformation and force
   $g_\alpha = (\zeta R_\alpha \omega)/2  $
   valid at lowest order in $a/L_\alpha$.
 
To produce a plot we 
  arbitrarily choose a phase shift  $\psi =\pi/2$  between the oscillators.
For a disk $\zeta = [16 a (\eta_s +\eta_c)]/3$. Moreover, we pose $R_A = R_B = R$ and
$g_A = g_B = g.$
So, 
  in the case where oscillations are driven by providing fixed 
  oscillation amplitude $R$
   we obtain 
    $
v_c 
 =
 \frac{\eta_s}{(\eta_s +\eta_c) r^4}
\frac{1}{\pi } a \omega
 l^2  R^2.
  $
 Instead, if oscillations are driven by providing fixed 
  force amplitude $ g$ we get 
$
v_c
 =
 \frac{\eta_s}{(\eta_s +\eta_c)^3 r^4}
\frac{1}{a  \omega}
\frac{9}{64 \pi}   l^2  g^2
  $.
  To distinguish between
  these two cases
 we   further pose
$ \Xi_g := 
\frac{1}{\eta^2_c}
\frac{1}{a \omega}
\frac{9}{64 \pi}   l^2  g^2 $  
and
$  \Xi_R := \frac{1}{\pi } a \omega
 l^2  R^2 $ 
  and 
 measure the migration
  speed respectively
   in  units of  $ \frac{{\Xi}_{R} }{ r^4}$ 
or  in  units of $ \frac{ {\Xi}_{g} }{ r^4}$,
   keeping $\eta_c$ constant and
 varying $\eta_s$
 see~Fig.(2)(a).
 A  different choice for the phase shift, ${\psi} \neq 0, \pi$, 
will introduce corrections associated to the  term $\cos {\psi}$ in  $\Xi$. 
As the amplitudes of oscillation must satisfy $R < l$,
 these corrections  do not affect the qualitative trend of the migration speed although they can  
  shift the value of substrate viscosity at  which  the migration is optimal, as well as the maximum speed.

\subsubsection{Approximated expression for the average migration speed for 
 elastic cells on viscous substrates}
  \label{sec:elastic-cell3}
 
In this case~\eq{eq:cdot-elast-cell}
depends on $\m v(\m x)$  given by~\eq{eq:v-elast-visc} and
  contains the sum of two terms.
 The first term  resembles what was found for the viscous cell lying on the viscous substrate.
The second term describes the contribution of the elastic interactions.
It is important to note that such a term
 contains an exact time-derivative, of periodic functions of period $T = \frac{2\pi}{\omega}$.
Thanks to that, 
the average over one period gives
$\int^T_0 dt \partial_t [ \bm{G}^{(e)} \cdot \m F ] = \bm{G}^{(e)} \cdot \m F |_{t=T} - \bm{G}^{(e)} \cdot \m F |_{t=0} \equiv 0$.
Thus for calculating the average speed we can approximate
$\m v(\m x) \sim \bm{H}^{(e)} \cdot \m F $ in ~\eq{eq:v-elast-visc}.
 The multipole expansion
for the elastic cell is identical to the one obtained for the viscous cell,~\eq{eq:c-dot-inst} (and related~\eq{eq:c-dot-inst2}). 
The difference is that for the elastic cell    forces $ f_\alpha$ and deformations $ d_\alpha$ 
 satisfy a  relation which follows from~\eq{eq:ddot-elast-cell}. 
 Neglecting, as before, the effect of the interactions at this level,
 \eq{eq:ddot-elast-cell}  
  yields  
  a local relation between forces and displacements of the oscillators
 $\dot{d}_\alpha = -\frac{1}{\mathcal{T}}d_\alpha + \frac{2}{\zeta_s} f_\alpha  $
 where $\mathcal{T} := \frac{\eta_s}{\mu_c} $ is the timescale obtained combining the elastic modulus of the cell 
  and the viscosity of the substrate.
Note that  $\mu_c/\omega$ has the dimension of a viscosity, so $ \omega \mathcal{T} =  \frac{\omega \eta_s}{\mu_c} $
is equivalent to a viscosity ratio.
 We obtain different force-deformation 
  relations depending on the chosen prescription.
   In particular we note the parametrisation given in~\eq{eq:param-d-f} needs to be modified to include the additional timescale $\mathcal{T}$, see below.

{\it Assigning forces.}
 If we assign forces as $f_\alpha = -g_\alpha  \sin (\omega t + \phi_\alpha) $ then
  displacements are obtained solving the differential equation for $d_\alpha$, whose solution
at the steady state is
 $d_\alpha(t) \sim -\frac{2}{\zeta_s} g_\alpha  \frac{\mathcal{T}}{1 +\omega^2 \mathcal{T}^2}[ \sin( \omega t+\phi_\alpha) -\omega\mathcal{T} \cos( \omega t+\phi_\alpha)  ]$.
As for a viscous cell, the dipole term does not contribute to the mean speed while the quadrupole does.
  Posing again $g_\alpha = g$ and $f_\alpha = f$ for $\alpha = A, B$ the result   is
\begin{align}
v_c(\psi)=    \frac{3 \eta_s g^2 \mathcal{T}^2 l^2  }{\eta^2_s +\frac{\mu^2_c}{\omega^2}  } 
\frac{\omega  [ (1 + \omega^2 \mathcal{T}^2) +(\frac{g \mathcal{T} }{ l \zeta_s })^2 \cos \psi ] \sin \psi}{ 4\pi \zeta_s r^4  (1+\omega^2 \mathcal{T}^2)^2 } 
\label{eq:vc-elast-prescr-force}
\end{align}

Note that the term $(\frac{g \mathcal{T} }{ \zeta_s })  $ has the dimension of length, and describes the oscillation amplitude  $d_\alpha$. Therefore it must satisfy $ (\frac{ g \mathcal{T} }{ \zeta_s }) <  l $.
Studying the behaviour as a function of $\eta_s$ we recover the non-monotonic trend
found for the case of fluid cell,  
peaked around $\eta_s \sim \frac{\mu_c}{\omega}$. 
It is interesting to note that
this migration speed is also a non-monotonic function of $\omega$, peaked around $\omega \sim \frac{2}{\mathcal{T}}$
(weakly sensitive to the values of $\psi$ and to the ratio $(\frac{ g \mathcal{T} }{ \zeta_s l})$).
This result suggests a cell may adapt its oscillation
 frequency to  the environment in order
  to optimise its speed.
  To produce a plot as a function of the frequency, in this case we consider the expression of $v_c$
   given by \eq{eq:vc-elast-prescr-force} for the value $\psi =\pi/2$ as before,  obtaining
  $ v_c =  \frac{3}{4 \pi}   \frac{l^2  g }{ \eta_s r^4} 
(\frac{ g }{\xi_c}) 
\frac{\omega^3 \mathcal{T}^3  }{(1+\omega^2 \mathcal{T}^2)^2}. 
$ 
\fig{fig:trend}(b) of the main text shows $ v_c $ in unit of the velocity $ \frac{3}{4 \pi}   \frac{l^2  g }{ \eta_s r^4} 
(\frac{ g }{\xi_c}) $ measuring $\omega$ in units of $ \mathcal{T}^{-1}$.

The physical reason beyond  the existence of an optimal
 frequency in the case of prescribed forces can be understood using a simple argument~\cite{LE+10}.
 The propulsion (and  pumping of fluid)
  are proportional to terms that contain product of forces, $f_\alpha$,
  and deformations, $d_\alpha$,  of the oscillators. 
  Imposing the force,
 in presence of viscoelasticity,
  the deformations  take place on typical timescale $\mathcal{T}$. Then:
 (i) 
 for sufficiently high frequencies, $ T \ll \mathcal{T}$, 
 the  oscillator cannot    reach the  maximum amplitude of oscillations.  
  In this regime the allowed amplitude of deformation increases at increasing $T$ (i.e. at decreasing $\omega)$;
  (ii) on the contrary
for sufficiently small frequencies, $ T \gg \mathcal{T}$, 
  the oscillator  can  
  reach the maximum amplitude available. So   in this regime
   the  amplitude of deformation does not vary at increasing $T$. 
 To calculate the average speed 
  we must divide $d_\alpha$ 
   by $T$.   The resulting function  in regime (i) can increase at increasing $T$ 
  while in regime  (ii) only decreases at increasing $T$ (behaving as $\sim \textrm{constant}/ T$). 
  The optimal frequency lies at the crossover between these two regions.
  The situation is different  when deformations are prescribed functions.

{\it Assigning deformations.}
If we assign deformations as $d_\alpha(t) = R \cos(\omega t +\phi_\alpha)$
The forces  are also prescribed functions of time, 
 directly related to the deformations via  
$ f_\alpha = \frac{\zeta_s}{2}[ \dot{d}_\alpha +\frac{d_\alpha}{\mathcal{T}} ]$
which yields
$ f_\alpha(t) = R \frac{\zeta_s}{2\mathcal{T}}[ - \omega \mathcal{T} \sin(\omega t +\phi_\alpha)  +  \cos(\omega t +\phi_\alpha) ]$.
Also here only the quadrupole contributes to the average as 
\begin{equation}
v_c(\psi) = \frac{3}{16 \pi}    \frac{\eta_s l^2 \omega \mathcal{T} }{\eta^2_s +\frac{\mu^2_c}{\omega^2} } 
(\frac{ R^2 \zeta_s }{ r^4 \mathcal{T}  }) 
[ 1  + \frac{ R^2}{4 l^2} \cos \psi ] \sin \psi. 
\label{eq:vc-elast-prescr-disp}
 \end{equation}
Interestingly, also this  expression has  a maximum as a function of the substrate viscosity but
 only  monotonic trend  as a function of $\omega$.
\fig{fig:trend}(b) of the main text  shows the velocity as a function of the frequency for fixed deformation. 
In this case we consider the expression of $v_c$
   given by \eq{eq:vc-elast-prescr-disp} for the value $\psi =\pi/2$ as before,  obtaining
 $v_c= \frac{1}{ \pi}    \frac{\omega^3 \mathcal{T}^3 }{1 + \omega^2 \mathcal{T}^2 } \frac{ R^2 l^2}{ r^4}    \frac{a}{ \mathcal{T}} $.
  We plot $ v_c $ in units of the velocity $\frac{1}{ \pi} \frac{ R^2 l^2}{ r^4}   
\frac{a}{ \mathcal{T}}  $ measuring $\omega$ in units of $ \mathcal{T}^{-1}$.

\bibliographystyle{apsrev4-1}
\bibliography{notes-letter}

\end{document}